\documentclass[10pt, journal]{IEEEtran}
\usepackage{float}
\usepackage[caption = false]{subfig}
\usepackage{array}
\usepackage{amsmath,amssymb,amsfonts}
\usepackage{algorithmic}
\usepackage{graphicx}
\usepackage{empheq}

\newcommand{\mc}[1]{\mathcal{#1}}
\newcommand{\mbb}[1]{\mathbb{#1}}
\newcommand{\mr}[1]{\mathrm{#1}}
\newcommand{\m}{\boldsymbol}
\newcommand{\bmat}[1]{\begin{bmatrix} #1 \end{bmatrix}}
\newcommand*\widefbox[1]{\fbox{\hspace{0em}#1\hspace{0em}}}

\newtheorem{asmp}{Assumption}
\DeclarePairedDelimiter\norm{\lVert}{\rVert}%

\DeclareMathOperator*{\minimize}{minimize}

\DeclareMathOperator*{\subjectto}{subject\, to}

\usepackage[colorlinks = true, linkcolor = blue, urlcolor  = blue, citecolor = blue]{hyperref}

\usepackage{lettrine}

\title{\Large \textbf{\textit{Robust Dynamic State Estimation of Multi-Machine Power Networks with Solar Farms and Dynamics Loads}}}
\author{Muhammad Nadeem and Ahmad F. Taha \vspace{-0.8cm}
\thanks{This work is supported by National Science Foundation under Grants 2152450 and 2151571. The authors are with  the Civil and Environmental Engineering Department, Vanderbilt University, 2201 West End Ave, Nashville, Tennessee 37235. {\small muhammad.nadeem@vanderbilt.edu, ahmad.taha@vanderbilt.edu}}%
}

\begin{document}

\maketitle

\begin{abstract}
Conventional state estimation routines of electrical grids are mainly reliant on dynamic models of fossil fuel-based resources. These models commonly contain differential equations describing synchronous generator models and algebraic equations modeling power flow/balance equations. Fuel-free power systems that are driven by inertia-less renewable energy resources will hence require new models and upgraded estimation routines. To that end, in this paper we propose a robust estimator for an interconnected model of power networks comprised of a comprehensive ninth order synchronous generator model, advanced power electronics-based models for photovoltaic (PV) power plants, constant power loads, constant impedance loads, and motor loads. The presented state estimator design is based on Lyapunov stability criteria for nonlinear differential algebraic equation (DAE) models and is posed as a convex semi-definite optimization problem. Thorough simulations studies have been carried out on IEEE-39 bus test system to showcase the robustness of the proposed estimator against unknown uncertainty from load demand and solar irradiance.
\end{abstract}

\section{Introduction and Paper Contributions}
\lettrine{P}{ower}  systems are moving rapidly toward a greener and more sustainable future through the widespread installation of renewable energy resources (RERs). 
The intermittent and volatile nature of RERs has created challenges in modern power systems and made the secure and reliable operation of the grid more difficult. The solution to these problems can be found in dynamic state estimation (DSE) for the next generation of smart power grids. In particular DSE can provide accurate realtime estimates of the physical states of electrical grid which can be highly beneficial for the realtime feedback control and health monitoring of electric grid \cite{Liu2021TPWRS}.

In this regard, thorough research has been carried out in the past two decades in power system DSE mainly focusing on estimating generator internal states (rotor angle and frequency) and algebraic states (voltage and current phasors). All of the current DSE algorithms can be divided into two broad categories: (1) deterministic observers and (2) Kalman filters (also known as stochastic estimators).
In deterministic observers the basic idea is as follows: First, a dynamical model for the error dynamics is computed (this model explains the evolution of error between original and estimated state variables), then observer gain is designed based on Lyapunov stability notion such that it drives the error model near zero \cite{SebastianITPWRS2020, SebastianITCNS2021, Jin2018MultiplierbasedOD}.

On the other hand Kalman filters (KF) are iterative and they primarily exploit the statistical properties (such as normal distribution) of the noise/disturbances to minimize their impact on the performance of DSE \cite{GhahremaniITPWRS2011,ZhaoITWPRS2018,ZhouITPWRS2013}.

A summary of all the techniques used for power system DSE can be found in recent survey paper \cite{Liu2021TPWRS}, while a thorough comparative study between different derivatives of Kalman filter can be found in \cite{ZhouITSG2015}. Similarly, a comparative analysis between stochastic estimators and deterministic observers can also be seen in \cite{JunjianIEEE_ACCCESS2018}. An extensive summary of these methods is beyond the scope of this paper.

In the current literature, the DSE algorithms are primarily based on a simple power system model capturing only dynamical model of generators and few algebraic constraints showcasing power balance equations. Thus the monitoring scope of DSE is mainly limited to the estimation of algebraic states and generator internal states. However, with the rapid deployment of solar PV power plants and dynamic loads in the electrical grid these traditional power systems models are gradually becoming obsolete and out of context \cite{Liu2021TPWRS}. Hence to achieve grid-wise situational awareness in the future power systems it is important to model solar PV power plants and dynamic loads in the nonlinear descriptor (NL-D) model of power system while performing DSE. 

Unfortunately, the literature on state estimation of multi-machine power network models with synchronous machines, solar PV power plants, and motor loads while considering PMU measurements models and algebraic constraints is virtually non-existent. In the recent study \cite{FangITSG2021} DSE has been extended and state estimation for a detailed PV power plant model has been reported, however, the study is limited to distribution network only and does not estimate algebraic variables and generators' internal states.

Based on the above discussion the contributions of this paper are as follows:
\begin{itemize}
	\item We propose DSE technique for a holistic model of a power system having ninth order synchronous generator model, advanced power electronics-based models of solar farms, motor loads, constant power loads, and constant impedance loads. The proposed DSE algorithm can simultaneously estimate all the states of the power system that include dynamic states of solar farm, loads and algebraic states. 
	\item To design a robust estimator that can handle load fluctuations and uncertainty from RERs, we propose an $H_{\infty}$ based state estimation technique. The main advantage of $H_{\infty}$ based state estimator over Kalman filters is that no prior knowledge about the statistic of uncertainty is required. 
	\item We showcase the performance of the proposed estimator on IEEE case39 test power system which is widely used for DSE studies in power systems.
\end{itemize}

The rest of the paper is organized as follows: Section \ref{sec:NL-D power system model} summarizes the NL-D model; Section \ref{section:robust_obs_design} discuses the proposed estimator design; Section \ref{sec:case studies} showcases simulation studies and the paper is concluded in Section \ref{sec:conclusion}.
\vspace{-0.2cm}
\section{Power Network Model}\label{sec:NL-D power system model}
We consider a graphical model of the electrical grid with $N$ number of buses, $\mathcal{N} = \left\lbrace 1,...,N\right\rbrace$ set of nodes and $\mathcal{E} \subseteq \mathcal{N}\times\mathcal{N}$ set of transmission lines. The overall power system is assumed to have: $G$ number of synchronous machines, $J$ number of PV power plants, and $L_p$, $L_z$, and $K$ number of constant power, constant impedance, and motor loads respectively. Notice that $\mathcal{N} = \mathcal{G}\cup \mathcal{R}\cup\mathcal{L}$, where $\mathcal{G}, \mathcal{R}$, and $\mathcal{L}$ denote a set of buses containing synchronous machines, renewables, and loads respectively. To that end, we model the interconnected model of power systems using a set of differential-algebraic equations as follows:
\begin{subequations}~\label{equ:PSModel}
	\begin{align}
		\vspace{-0.5cm}
		\textit{system dynamics:} \;\;\;	\dot{\m x}(t) &= \m f(\m x_d,\m x_a, \m u, \m w)  ~\label{equ:PSModel-a} \\
		\textit{algebraic constraints:} \;\;\;	\m 0 &= \m h(\m x_d,\m x_a, \m u, \m w) ~\label{equ:PSModel-b}
	\end{align}
 \end{subequations}
where $\m x_d \in \mathbb{R}^{n_d}$ denotes differential variables, $\m x_a \in \mathbb{R}^{n_a}$ denotes algebraic variables, $\m w \in \mathbb{R}^{n_w}$ models the exogenous disturbances to the system, and $\m{u} \in \mathbb{R}^{n_u}$ models a set of commands that are used to drive the grid to a desirable state. 

In \eqref{equ:PSModel} we model $\m x_a$ as:
\vspace{-0.2cm}
\begin{align}\label{eq:x_a}
	\m x_a(t):=\m x_a = \bmat{\m V_{Re}^\top&\m V_{Im}^\top&\m I_{Re}^\top&\m I_{Im}^\top}^\top \vspace{-0.3cm}
\end{align}
where $\m V_{Re}\hspace{-0.01cm}= \hspace{-0.01cm}\{V_{Re_i}\}_{i\in \mc{N}}\hspace{-0.01cm}, \m V_{Im}\hspace{-0.01cm}=\hspace{-0.01cm} \hspace{-0.01cm}\{V_{Im_i}\}_{i\in \mc{N}}\hspace{-0.01cm}, \m I_{Re}\hspace{-0.01cm}=\hspace{-0.01cm} \hspace{-0.01cm}\{I_{Re_i}\}_{i\in \mc{N}}\hspace{-0.01cm}$ and $\m I_{Im}\hspace{-0.01cm}= \hspace{-0.01cm}\{I_{Im_i}\}_{i\in \mc{N}}\hspace{-0.01cm}$ denotes the real and imaginary part of voltage and current phasors respectively.

The vector $\m{u}$ contains reference inputs to the generators and PV plant and is modeled as:
\vspace{-0.2cm}
\begin{align}
	\m u = \bmat{\m u_G^\top & \m u_R^\top}^\top
\end{align}
where $\m u_G$ contains voltage reference set points $\m V_{ref}$ and steam/hydro valve reference positions $\m P_{v_{ref}}$ for the synchronous machines, such that $ \m u_G = \bmat{\m V_{ref}^\top & \m P_{v_{ref}}^\top}$. Similarly $ \m u_R = \bmat{\m V_{ref}^\top & \m P_{ref}^\top}$, where $\m V_{ref}$ is the voltage reference and $\m P_{ref}$ is the power reference set points for PV power plants.

Similarly, vector $\m w$ in Eq. \eqref{equ:PSModel} is expressed as follows:
\vspace{-0.2cm}
\begin{align}
	\m w = \bmat{\m P_{d}^\top & \m I_{r}^\top}^\top
\end{align} 
where $ \m P_d$ is the disturbance in load demand and $ \m I_r$ is the disturbance in the irradiance from the sun.
Furthermore in \eqref{equ:PSModel} we model $\m x_d$ as:
\vspace{-0.15cm}
\begin{align}\label{eq:x_d}
	\m x_d(t):=\m x_d = \bmat{\m x_G^\top&\m x_R^\top&\m x_m^\top}^\top
	\vspace{-0.1cm}
\end{align} 
where $\m x_G$ denote states of generators, $\m x_R$ denote dynamic states of PV plant, and  $\m x_m$ represents dynamic states of motor load. To that end, we model synchronous machines using a comprehensive ninth order generator model, thus $\m x_G$ can be given as follows \cite{sauer2017power,HugoITPWRS2019}:
\vspace{-0.24cm}
\begin{align}
	\m x_G = \bmat{\m e_{dq}^\top\;\; \m \omega_{sg}^\top \;\; \m \delta_{sg}^\top\;\;\m T_{m}^\top\;\;\m P_{v}^\top\;\;\m E_{fd}^\top\;\;\m v_{a}^\top\;\;\m r_{f}^\top}^{\top}\in \mbb{R}^{9G}
\end{align}
where $\m e_{dq} = [\m e_{d}^\top\,\,\m e_{q}^\top]$ denotes transient voltages along q- and d-axis, $\m \omega_{sg}$ is the rotor speed, $\m \delta_G$ is the rotor angle, $\m T_{m}$ represent prime mover torque, $\m P_{v}$ is the valve position, $\m E_{fd}$ denotes field voltage, $\m v_{a}$ represents amplifier voltage,  and $\m r_{f}$ denotes stabilizer output. 

The models for the PV power plant has been obtained 
from \cite{HugoITPWRS2019} and thus the state vector $\m x_R$ can be expressed as follows:
\begin{align}
\small \m x_R\hspace{-0.01cm} =\hspace{-0.01cm} \bmat{\m i_{dq_f}^\top\;\; \m v_{dq_c}^\top \;\;\m E_{dc}^\top\;\; \m P_{e}^\top\;\; \m Q_{e}^\top\;\;\m \delta_{inv}^\top \;\;\m z_{dq_0}^\top \;\;\m z_{dq_f}^\top}^{\top}\hspace{-0.1cm} \in\mbb{R}^{12J}
\end{align}
where $\m{i}_{dq,f}$ denotes the \emph{dq}-axis current at the terminals of the inverter of the PV plant, $\m{v}_{dq,c}$ is the \emph{dq}-axis voltage across the AC capacitor, $\m{E}_{dc}$ the energy stored in the dc-link capacitor of the PV plant, $\m{{P}_{e}}$ and $\m{{Q}_{e}}$ are the filtered real and reactive power at the terminals of the PV power plant, $\m{\delta_{inv}}$ is the relative angle of the PV power plant, $\m{z}_{dq,o}$ and $\m{z}_{dq,f}$ are the states of the voltage and current regulators used in the grid-forming controller of the PV plant.
Readers are referred to \cite{HugoITPWRS2019} for further details and complete in-depth description about the PV plant model.

The dynamics of the motor loads are detailed as \cite{krause2013}:
\vspace{-0.2cm}
\begin{equation}
	\dot{ \omega}_{mot_k} = \frac{1}{2H_{m_k}}(T_{e_k} - T_{m_k}) \label{eq:omegam}
\end{equation}
where $H_{m_k}$ is the motor inertia constant and $T_{e_k}$, $T_{m_k}$ denote electromagnetic and mechanical  torques in the $k$-th motor \cite[p. 244]{krause2013}.  Thus $\m x_m = [\m w_{mot_k}]$. 

To take into account the topological effect of power systems, the power flow/balance or the current balance equations need to considered. These can be expressed as follows~\cite{sauer2017power}:
\begin{gather}
	\underbrace{\begin{bmatrix}
		\m{\widetilde{I}}_{G} \\ \m{\widetilde{I}}_{R} \\ \m{\widetilde{I}}_{L}
	\end{bmatrix}}_{\m I}
	-
	\underbrace{\begin{bmatrix}
		\m{Y}_{GG} & \m{Y}_{GR} & \m{Y}_{GL} \\
		\m{Y}_{RG} & \m{Y}_{RR} & \m{Y}_{RL} \\
		\m{Y}_{LG} & \m{Y}_{LR} & \m{Y}_{LL} 
	\end{bmatrix}}_{\m Y}
	\underbrace{\begin{bmatrix}
		\m{\widetilde{V}}_{G} \\ \m{\widetilde{V}}_{R} \\ \m{\widetilde{V}}_{L} \\
	\end{bmatrix}}_{\m V} =  \m{0} \label{eq:transalgebraic}
\end{gather}
where $\m I$ is the net injected current vector, $\m{Y}$ is the admittance matrix, and $\m V$ is the bus voltage vector. In \eqref{eq:transalgebraic}, $\m{\widetilde{I}}_{G}\hspace{-0.13cm}= \hspace{-0.13cm}\{I_{Re_i}\}_{i\in \mc{G}}\hspace{-0.05cm}+\hspace{-0.05cm}j \{I_{Im_i}\}_{i\in \mc{G}}\hspace{-0.05cm}$ denotes phasor currents injected by synchronous generators and $\m{\widetilde{V}}_{G} \hspace{-0.06cm}= \hspace{-0.06cm}\{V_{Re_i}\}_{i\in \mc{G}}\hspace{-0.05cm}+\hspace{-0.05cm} j\{V_{Im_i}\}_{i\in \mc{G}}\hspace{-0.05cm}$ represents voltage phasors at the terminal of generator buses. Similarly $\m{\widetilde{V}}_{R}\hspace{-0.13cm}$ , $\m{\widetilde{V}}_{L}$,  $\m{\widetilde{I}}_{R}$, and $\m{\widetilde{I}}_{L}$ denotes voltage and current phasors of all loads and PV power plants. 

To that end, by considering (\ref{eq:x_a})--(\ref{eq:transalgebraic}) and incorporating the associated dynamics then the overall interconnected model of power systems can be represented in a following compact form:
\vspace{-0.1cm}
\begin{subequations}\label{eq:final_DAE}
	\begin{empheq}[box=\widefbox]{align}
\textbf{NL-D:}\;\;\; 		\m E\dot{{\m x}} &= {\m A}{\m x} +  {\m f}\left({\m x},{\m u},{\m w} \right) + {\m B} {{\m u} } + {\m B}_w \m w \\
		\m y &=   \m C \m x + \m v \label{eq:final_DAE2}
	\end{empheq}
\end{subequations}
where $\m x = \bmat{\m x_d^\top & \m x_a^\top}^\top \in\mbb{R}^{n}$ denotes the overall state vector; $\m E$ encodes algebraic constraints with rows of zeros; $\m C$ maps state vector $\m x$ to what PMUs usually measures (i.e., current and voltage phasors) and $\m v\in\mbb{R}^{p}$ denotes measurement noise on PMU measurements $\m y \in \mathbb{R}^{p}$.  Since vector $\m x_a$ contains voltage and current phasors, we define $\m C$ as: $\m C=\bmat{\m O&\tilde{\m C}}$, where $\tilde{\m C}$ is a diagonal binary matrix with ones only at those locations where PMUs are connected and voltage and current phasors are measured. The other state-space matrices $\m A, \m B$, and $\m B_w$ are obtained by capturing the linear components of the SLS-DAE model. The vector-valued function $\m f(\cdot)$ captures the encompassed nonlinearities in the dynamics.
\section{Estimator for NL-D Model of Power Systems}\label{section:robust_obs_design}
In this section, we propose a Luenberger type observer design for the NL-D model depicted in \eqref{eq:final_DAE}. First, we focus on modeling the uncertainties in loads and renewables and present a technique to parameterize the nonlinear function $\m f(.)$ in the power system dynamics.
\vspace{-0.15cm}
\subsection{Modeling uncertainty from loads and renewables}
In Eq. \eqref{eq:final_DAE} the vector $\m w$ encapsulates load demand and irradiance, note that both of these quantities are time-varying and fluctuating. To that end, herein we assume that hour- or minute-ahead predictions of loads and irradiance are available while the fluctuation/disturbances in these quantities are unknown. Notice that this is realistic as power systems operators record and publish these quantities on a daily basis (see daily hour- and minute-ahead predictions of load and renewables published by California independent system operator \cite{CAISO}). However, the prediction may be inaccurate. In particular, high fidelity estimate of RERs are difficult to obtain. Accordingly, we can write $\m w = \bar{\m w} + \Delta \m w$, where $\bar{\m w}$ is the predicted/known values and  $\Delta \m w$ defines all the disturbances/uncertainties in these quantities. The goal of the estimator is to provide accurate state estimation results under unknown uncertainty $\Delta \m w$. To that end, the NL-D model \eqref{eq:final_DAE} can be rewritten as follows:
\begin{subequations}\label{eq:final DAE}
		\vspace{-0.2cm}
	\begin{align}
		\begin{split}
			\m E\dot{{\m x}}\hspace{-0.01cm}&=\hspace{-0.01cm}{\m A}{\m x}\hspace{-0.008cm}\hspace{-0.01cm} + \hspace{-0.01cm} {\m f}\left({\m x},{\m u},\m w \right) \hspace{-0.01cm}+ \hspace{-0.01cm}{\m B} {{\m u} }\hspace{-0.01cm} +\hspace{-0.01cm} {\m B}_w \bar{\m w}\hspace{-0.01cm} +\hspace{-0.01cm} {\m B}_w \Delta \m w  
		\end{split}\label{eq:final DAE1}\\
		\begin{split}
			\m y &=   \m C \m x+\m v. 
		\end{split}\label{eq:final DAE2}
	\end{align}
\end{subequations}
The main reason for splitting the vector $\m w$ into known/predicted and unknown parts is because later on in the estimator dynamics to achieve robust performance the estimator only has access to the predicted value of $\m w$ and disturbances $\Delta \m w$ is kept unknown to the estimator.

In the following section we discuss Lipschitz continuity condition to classify the nonlinear vector valued function $\m f(.)$ in the system dynamics.

\subsection{Parameterizing nonlinearities in the NL-D model}
To synthesize robust observers it is crucial to identify and parameterize the nonlinearities in the dynamical system. 
Hence, to express the nonlinear function $\m f(.)$ in a better way in the NL-D model we assume that $\m f(.)$ is Lipschitz continuous, which means $\m f(.)$ is differentiable everywhere and the rate of change is bounded above by a real number. Smallest such real number is called Lipschitz constant. With that in mind we assume $\m x\in \mathcal{X}$, where $\mathcal{X}$ define the operating region of state vector $\m x$, then the Lipschitz bounding condition for $\m f(.)$ can be written as follows:
\begin{align}\label{eq:lipshitz}
	\norm{\m f(\m x, \m u, \m w) - \m f(\hat{\m x}, \m u, \m w)}_2 \leq \alpha\norm {(\m x - \hat{\m x} )}_2
\end{align}
where $\alpha$ is the Lipschitz constant. In the case of power systems the Lipschitz constant can be assumed based on operator knowledge and by looking at the overall size/topology of the grid or it can also be computed using much more sophisticated ways as shown in \cite{SebastianACC2019,NugrohoITAC2021}. Also, since the Lipschitz constant provides an operational bound on grid state variables, its value can be varied to consider anomalies and deviations from steady state values and vice versa \cite{SebastianITPWRS2020}. Showcasing methods to compute Lipschitz constant is beyond the scope of this paper.
\vspace{-0.2cm}
\subsection{$H_{\infty}$ based state estimator design}
In this study we are designing a Luenberger type state estimator for NL-D model of power systems depicted in \eqref{eq:final DAE}. The overall estimator design is primarily based on Lyapunov stability criterion and $H_{\infty}$ estimation concept is used to achieve robust performance under unknown fluctuations/disturbances from loads and RERs.  

In state estimation literature, the $H_{\infty}$ concept was initially introduced in \cite{ShakedITAC1990} to synthesize a robust estimator for linear systems subject to unknown uncertainty. The main advantage of $H_{\infty}$ based state estimation is that no prior knowledge or assumptions are required about the statistical properties of uncertainty. In $H_{\infty}$ based state estimation the uncertainty is considered as a random bounded signal and then the observer is designed such that it ensures a particular $H_{\infty}$ performance for the error dynamics for all such bounded uncertainty, such that: $\norm{{\m e }}^2_{L_2}\leq\gamma^2\norm{{\Delta \m w }}^2_{L_2}$ with performance level $\gamma$. 

That being said, we now focus on presenting the structure of our observer design for the NL-D model of power system depicted in \eqref{eq:final DAE}. To begin, let $\hat{\m y}$ be the estimated outputs and $\hat{\m x}$ be the estimated states variables. Then Luenberger type state estimator for Eq. \eqref{eq:final DAE} can be expressed as follows:
\vspace{-0.1cm} 
\begin{subequations} \label{eq:obsr dynamics}
	\begin{align} 
		\begin{split}
			\m E\dot{\hat{\m x}}\hspace{-0.01cm} &= \hspace{-0.01cm}{\m A}{\hat{\m x}}\hspace{-0.01cm} +\hspace{-0.01cm} {\m f}\left({\hat{\m x}},{\m u},{\bar{\m w}}\right) +\m L(\m y -{\hat{\m y})}+ {\m B} {{\m u} }+ {\m B_w} {{\bar{\m w}}}
		\end{split} \label{eq:obsr dynamics1}\\
		\begin{split}
			\hat{\m y} &= \m C \hat{\m x}
		\end{split} \label{eq:obsr dynamics2}
	\end{align}
\end{subequations}
where $\m L \in \mbb{R}^{n\times p}$ is the estimator gain matrix. Even if the estimator starts from different initial conditions, using measurement provided by PMUs $\m y$, the gain matrix $\m L $ ensures the convergence of estimated states $\hat{\m x}$ to their true values $\hat{\m x}$. The main objective of this paper is to design an appropriate gain matrix $\m L $ such that robust state estimation can be achieved from the estimator dynamics \eqref{eq:obsr dynamics}. 

To that end, we define the error between actual and true values of state variables as $\m e = \m x - \hat{\m x}$. Then using \eqref{eq:final DAE}  and \eqref{eq:obsr dynamics} the model for the error dynamics can be computed as:
\vspace{-0.2cm}
\begin{align}\label{eq:error dynamics}
	\m E\dot{{\m e}} &= \m A_c\m e + \Delta \m f + \m B_w \Delta\m w
	\vspace{-0.2cm}
\end{align}
where $\m A_c = (\m A -\m L \m C)$ and $ \Delta \m f =\m f \left({\m x},{\m u}, {\m w}\right)-{\m f}\left({\hat{\m x}},{\m u},{\m w}\right)$. The primary objective of the estimator is to drive the differential equation of the error \eqref{eq:error dynamics} near zero under unknown uncertainty $\Delta \m w$. Before we proceed to the theory of designing the gain matrix $\m L$, we present key assumption that is critical in developing the proposed methodology.

\begin{asmp}\label{asmp:Asmption on nonlinearity}
	The effect of uncertainty $\Delta\m w $ on the norm of $\m f(.)$ is negligible, meaning: 
	\vspace{-0.2cm}
	\begin{align}
		\norm{\m f(\hat{\m x}, \m u, \bar{\m w})}_2 \approx \norm{\m f(\hat{\m x}, \m u, {\m w})}_2.
	\end{align}
\end{asmp}

Assumption \ref{asmp:Asmption on nonlinearity} is crucial in developing the theory of the observer design as it is required while applying the S-procedure lemma \cite{Slemma} in the derivation of estimator gain in a tractable way. Notice that Assumption \ref{asmp:Asmption on nonlinearity} is mild and holds in the case of power networks as the nonlinear function in the power systems models are commonly independent of uncertainties and are mostly dependent only on state vector, see \cite{ZhaoITPWRS2020} and references therein. To validate this on the NL-D model \eqref{eq:obsr dynamics} we used different IEEE test power systems such as WECC-9 bus system and IEEE-39 bus system and we note that by varying $\m P_d$ in $\m w$ the norm of $\m f(.)$ does not change while by adding uncertainty in $\m I_r$ the norm of $\m f(.)$ changes slightly, such as by decreasing the irradiance by $40\%$ the percentage change in $\norm{\m f(.)}_2$ is $0.09\hspace{-0.05cm}\times\hspace{-0.05cm}10^{-3}\%$. Hence Assumption \ref{asmp:Asmption on nonlinearity} is empirically satisfied. Proving Assumption \ref{asmp:Asmption on nonlinearity} theoretically is an interesting research question beyond the scope of this paper.

With that in mind, we can find necessary and sufficient conditions for the existence of observer gain matrix $\m L$. To that end, let us assume a Lyapunov candidate function $V(\m e)=\m e^\top \m E^\top \m P \m e $, where  $V:\mbb{R}^{n}\rightarrow \mbb{R}_+$, $\m P\in\mbb{R}^{n\times n}$, and $\m E^\top \m P = \m P^\top \m E \succeq 0$, then its derivative along the trajectories of Eq. \eqref{eq:error dynamics} yields: 
\begin{align*}
	{\dot V}(\m e) &= (\m{E}\dot{\m e})^\top \m P \m e+(\m P \m e)^\top(\m{E}\dot{\m e}).
\end{align*}
For any bounded disturbances $\Delta\m w$ the $H_{\infty}$ stability condition can be satisfied as: ${\dot V}(\m e) + \m e^\top\m e - \gamma^2 \Delta \m w^\top\Delta\m w < 0$, which can be written as:  $\m\Upsilon^\top\m\Theta_1\m\Upsilon<0$, where
\begin{align*}
	\m\Upsilon = \bmat{\m e\\\Delta \m f\\\Delta\m w}, \m\Theta_1=\hspace{-0.01cm}\bmat{\m A_c^\top \m  P\hspace{-0.01cm}+\hspace{-0.01cm}\m P^\top \m A_c\hspace{-0.01cm}+\hspace{-0.01cm}\m I  & \m P^\top & \m P^\top\m B_w\\ \m P & \m O &\m O \\
		\m{B}_w^\top\m{P}&\m O&-{\gamma}^2\m I}.
\end{align*}
Now from Lipschitz continuity assumption \eqref{eq:lipshitz} we obtain: $\Delta \m f^\top \Delta \m f-\alpha^2 \m e^\top \m e\leq 0$, which can be written as $\m\Upsilon^\top\m\Theta_2\m\Upsilon\leq0$, with $\m\Theta_2 = \mr{diag}\left( \bmat{-\alpha^2 \m I&\m I&\m O}\right)$,
where $\mr{diag}$ denotes a diagonal matrix. Now applying the S-procedure lemma \cite{Slemma}, then $\m\Theta_1- (\epsilon)\m\Theta_2\prec0$ for some scalar $\epsilon>0$, which is equivalent to:
\begin{align}\label{proof: LMI_int}
	\hspace{-0.01cm}\bmat{\m A_c^\top\m P\hspace{-0.01cm}+\hspace{-0.01cm}\m P^\top\m A_c\hspace{-0.01cm}+\hspace{-0.01cm}\m I+\epsilon\alpha^2\m I  & \m P^\top & \m P^\top\m B_w\\ \m P & -\epsilon\m I &\m O \\
		\m{B}_w^\top\m{P}&\m O&-{\gamma}^2\m I}\prec0.
\end{align}

Finally, by rewriting $\m P$ as $\m{P} = \m{XE}+{\m E^{\perp\top}}{\m Y}$ for some $\m Y\in \mbb{R}^{n_a\times n}$ and $\m X\in \mbb{S}^{n\times n}_{++}$, where $\m E^{\perp\top}$ is the orthogonal complement of $\m E$ \cite[ch. 2]{Xu} and defining $\m N= \m L^\top \m P\in\mbb{R}^{p\times n}$ then the necessary and sufficient condition for the existence of observer gain $\m L$ can be written in term of strict linear matrix inequality (LMI) as follows: 
\begin{align}\label{eq:LMI_Hinf}
	\hspace{-0.1cm}\bmat{ \m\Omega_{11} & * & * \\\m X\m E+\m E^{\perp\top}\m Y & -{\epsilon}\m I &* \\
		\m{B_w}^\top\m X\m E+\m{B_w}^\top\m E^{\perp\top}\m Y&\m O&-{\gamma}^2\m I} \prec 0 
\end{align}
where $\m\Omega_{11} $ is given as:\vspace{-0.1cm}
\begin{align*}
	\begin{split}
		\m\Omega_{11} = \m A^\top\m X\m E+\m A^\top\m E^{\perp\top}\m Y+ \m E^\top\m X^\top\m A+\m Y^\top\m E^{\perp}\m A -\\
		\m C^\top \m N-\m N^\top \m C +\epsilon \alpha^2 \m I.
	\end{split}
\end{align*}

If LMI \eqref{eq:LMI_Hinf} is solved and there exist matrices $\m N$, $\m Y$, $\m X $, and $\gamma$ and $\epsilon>0$, then the estimator gain can be retrieved as $\m L=\left( {\m N\m P^{-1}}\right) ^\top$.

Notice that, since $\gamma$ denotes performance level, then one can minimize $\gamma$ to achieve robust performance. Also, minimizing  $\norm{\m N}_2$ can provide estimator gain of reasonable magnitude. Furthermore, as our Lyapunov candidate function can be written as $V(e) =\m e^\top\m E^\top\m X\m E\m e$ then to achieve quick convergence of $\hat{\m x}$ to $\m x$ one can minimize the maximum eigenvalue of $\m E^\top\m X\m E$ which can be written as a convex SDP optimization problem: 
\vspace{-0.2cm}
\begin{align*}
	\mathbf{\left( P_1\right) }\;\minimize_{\kappa, \m X}  \;\;\;& a_3\kappa\;\;\;\; \mathrm{subject\;to}\,\,\;\;\; \kappa\m I-\m E^\top\m X\m E\succ 0, \kappa>0.
\end{align*}
Hence, the necessary and sufficient condition of computing $\m L$ given in Eq. \eqref{eq:LMI_Hinf} can be converted to a convex semidefinite optimization problem expressed as follows:
\vspace{-0.2cm}
\begin{align*}
	\mathbf{\left( P_2\right) }\;\minimize_{\kappa,{\epsilon},\gamma, \m N, \m X, \m Y}  \;\;\;& a_1\norm{\m N}_2+ a_2\gamma + a_3\kappa\\ \subjectto \;\;\; &  \kappa\m I-\m E^\top\m X\m E\succ 0,\; \mr{LMI}\; \eqref{eq:LMI_Hinf},\;\\&\m X \succ 0, \; \kappa>0,\;{\epsilon}> 0,\; \gamma > 0
\end{align*}
where $a_1$, $a_2$, $a_3$ $\in \mbb{R_{++}}$ are predefined weighting constants. Calculating estimator gain matrix $\m L$ by solving $\mathbf{P_2}$ always ensures that the performance of state error dynamics \eqref{eq:error dynamics} is bounded in  $H_{\infty}$ sense, such that $\norm{{\m e }}^2_{L_2}\leq\gamma^2\norm{{\Delta \m w }}^2_{L_2}$. In other words, solving $\mathbf{ P_2 }$ guarantees that $\norm{{\m e }}^2_{L_2}$ always lies in a tube of circle with center at the origin and having a radius of $\gamma^2\norm{{\m w }}^2_{L_2}$.

In the following section we present thorough numerical studies to assess the performance of the estimator under various uncertainties from loads and renewables.
\section{System Setup and Simulation Studies}\label{sec:case studies}
\begin{figure}[h]
	\vspace{-0.3cm}
	\centering
	\includegraphics[keepaspectratio,scale=0.4]{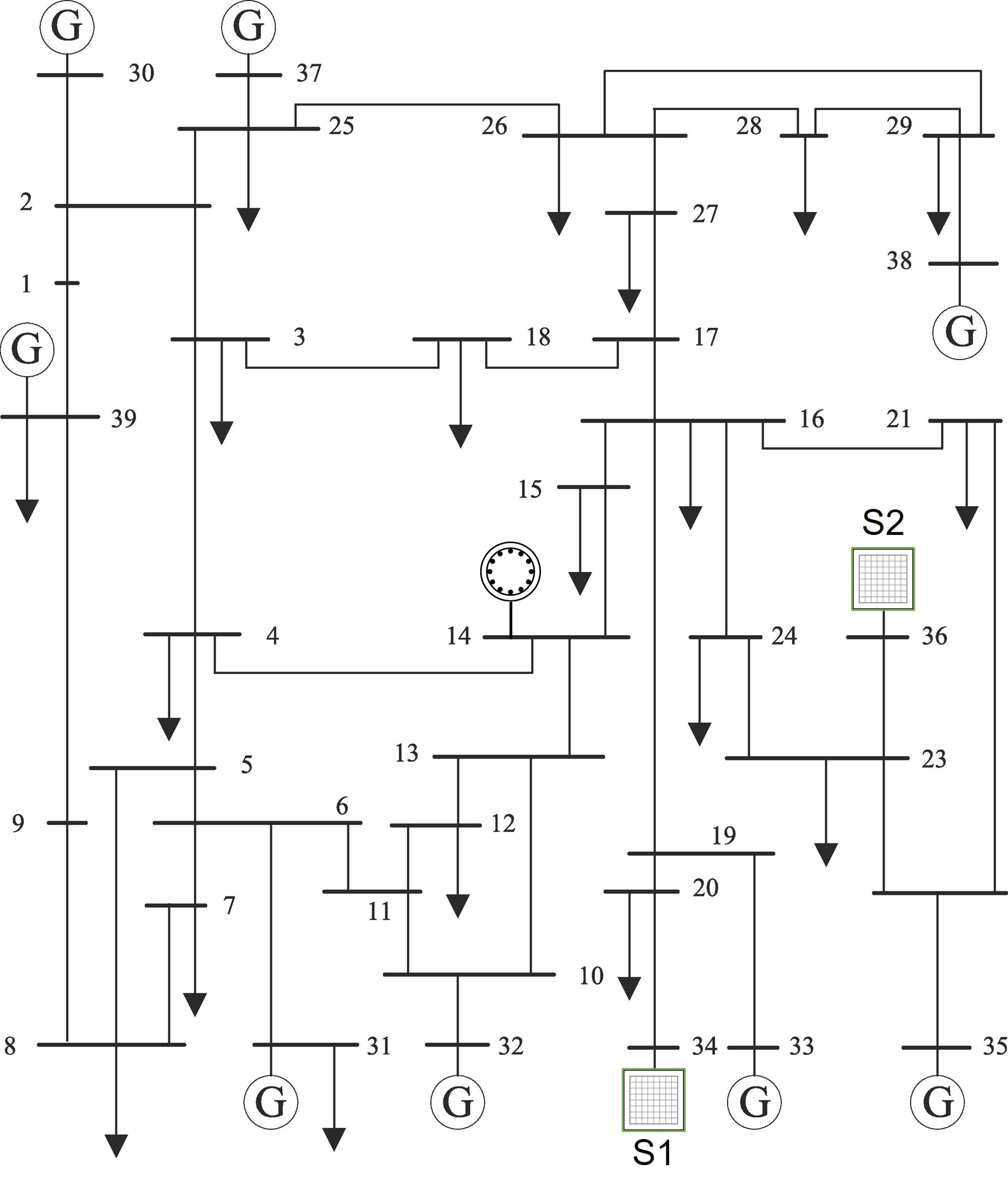}\vspace{-0.41cm}\caption{IEEE 39 bus system with  two PV power plants at Buses 34, 36 and a motor load at Bus 14.}\label{fig:ieee39}\vspace{-0.24cm}
\end{figure}

The proposed estimator has been tested on modified IEEE-39 bus system \cite{Hiskens}. This system consists of two PV power plants at Buses 34 and 36, a motor load at Bus 14, eight synchronous generators, constant power loads and constant impedance loads. The one-line diagram of this system is depicted in Fig. \ref{fig:ieee39}. All the parameters for the synchronous generator and its excitation systems are detailed in \cite{sauer2017power}, while the parameters of PV power plants and other complete in-depth description about the system model can be found in \cite{HugoITPWRS2019}.
We deploy 13 PMUs at Buses $[2,6,9,10,13,14,17,19,20,22,23,25,29]$ as given in \cite{ChakrabartiITPWRS2008} to ensure that the power system is observable. Each PMU at a bus is measuring the total current demanded/injected and voltage phasors of that bus. The dynamic states that we are estimating are $\m x_a$ and $\m x_d$ as given in \eqref{eq:x_a} and \eqref{eq:x_d}. The uncertain quantities for the estimator are assumed to be the overall load demand and irradiance from the sun. Moreover, for all the case studies Gaussian noise with diagonal covariance matrix and variance of $0.001^2$ has also been added to the PMU measurements.

Both the estimator and power system NL-D models are simulated using MATLAB  differential algebraic equations solver \texttt{ode15s}, whereas all the SDPs are solved in YALMIP \cite{LofbergICRA2004} with MOSEK \cite{Andersen2000} as optimization solver. The initial conditions for the power system are computed using power flow solutions, while the estimator is initialized from random initial conditions having $10\%$ maximum variation from the steady state values of power system. Through out the simulations we consider weighting constant in $\mathbf{P_2}$ as $a_1 = a_2 = a_3 = 1$, power system base as $S_b = 100\mr{MVA}$, and $\omega_0 = 2\pi60$ $\mr{{rad/sec}}$. 
\vspace{-0.5cm}

\subsection{Case 1: Estimation under demand disturbance}

In power systems the overall load demand is usually fluctuating and the exact estimate of the load demand of any power network is difficult to predict. To that end, herein we demonstrate the performance of the proposed estimator in performing DSE under unknown disturbances in power demand. To showcase the robustness of the estimator we provide only steady state values of loads to the estimator and the actual transient/fluctuating values are kept unknown. 

\begin{figure}[h]
	\centering
 \vspace{-0.5cm}
	\subfloat[]{\includegraphics[keepaspectratio=true,scale=0.42]{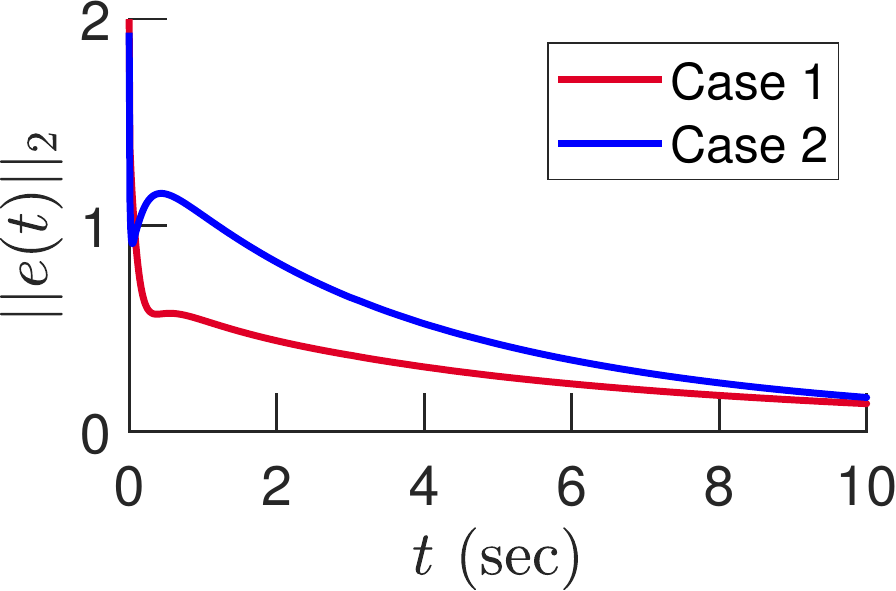}}{}\hspace{-0.01cm}
	\subfloat[]{\includegraphics[keepaspectratio=true,scale=0.42]{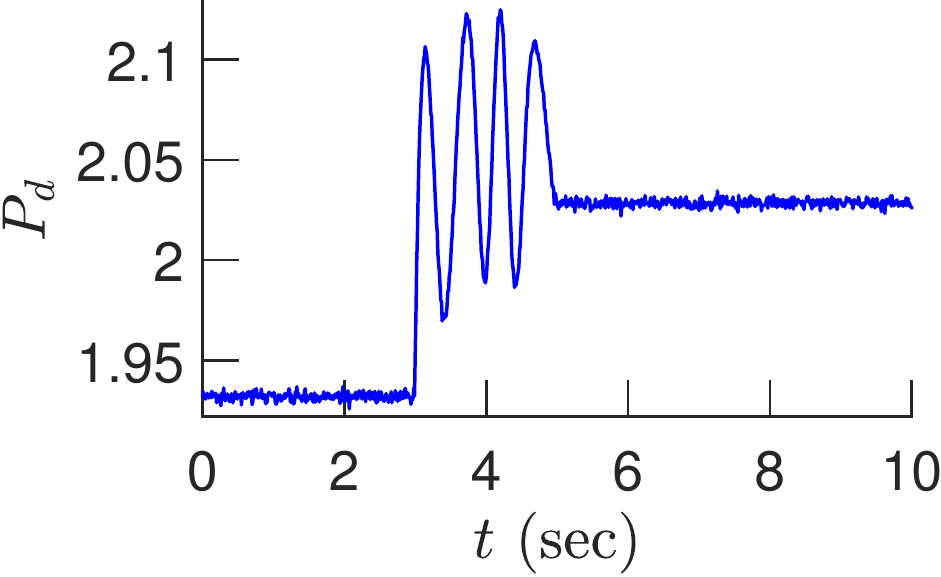}}{}\hspace{-0.01cm}\vspace{-0.1cm}\caption{(a) Estimation error norm for Case 1, Case 2 and (b) Variation in real power demand at Bus 3 and 4.}
	\label{fig:Pd and error}
\end{figure}

With that in mind, to generate random fluctuations in the constant power loads we provide step disturbances at $t = 3$s to the loads connected at Buses $[7,8,12,14,15,16,18,20]$, while the loads connected at Buses $3$ and $4$  are assumed to be changing sinusoidally with some Gaussian noise as shown in Fig. \ref{fig:Pd and error}.
\begin{figure}
	\centering
	\subfloat{\includegraphics[keepaspectratio=true,scale=0.42]{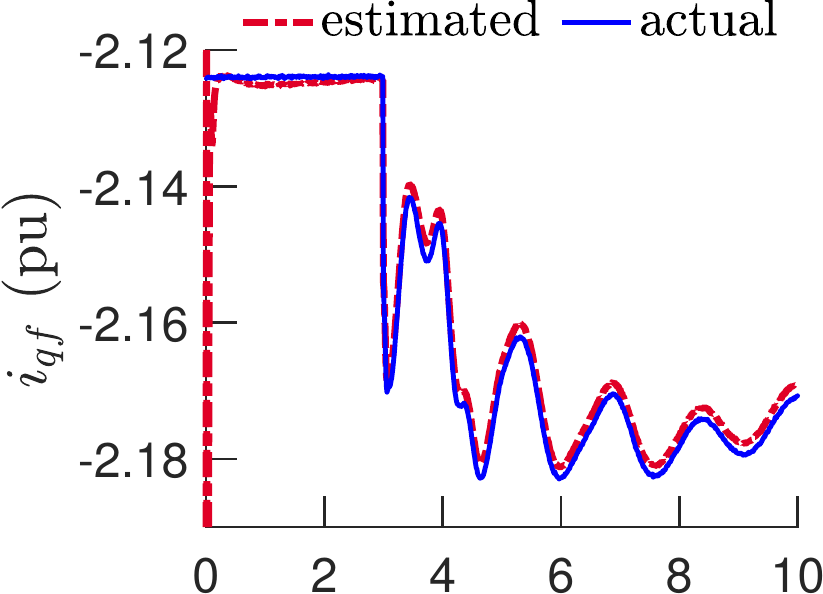}}{}{}
	\subfloat{\includegraphics[keepaspectratio=true,scale=0.42]{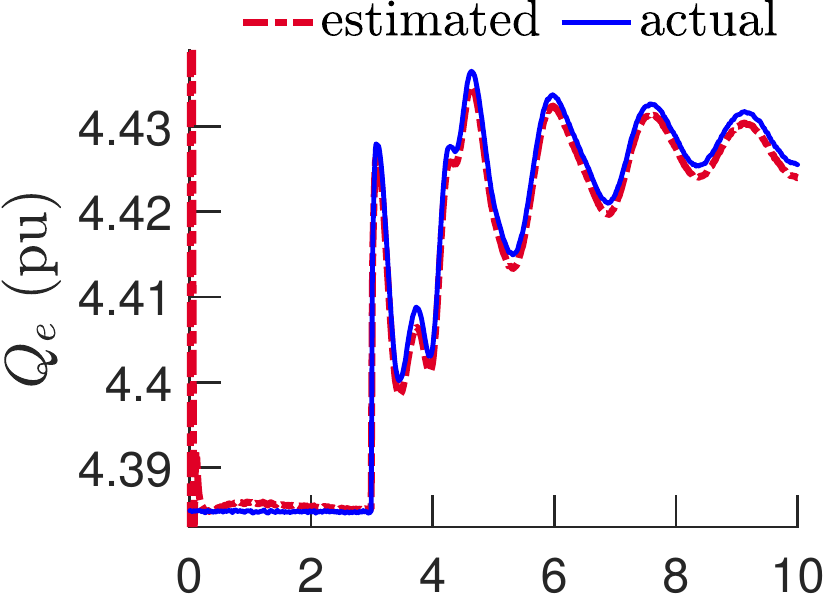}}{}{}\vspace{0.19cm}\hspace{-0.15cm}
	\subfloat{\includegraphics[keepaspectratio=true,scale=0.42]{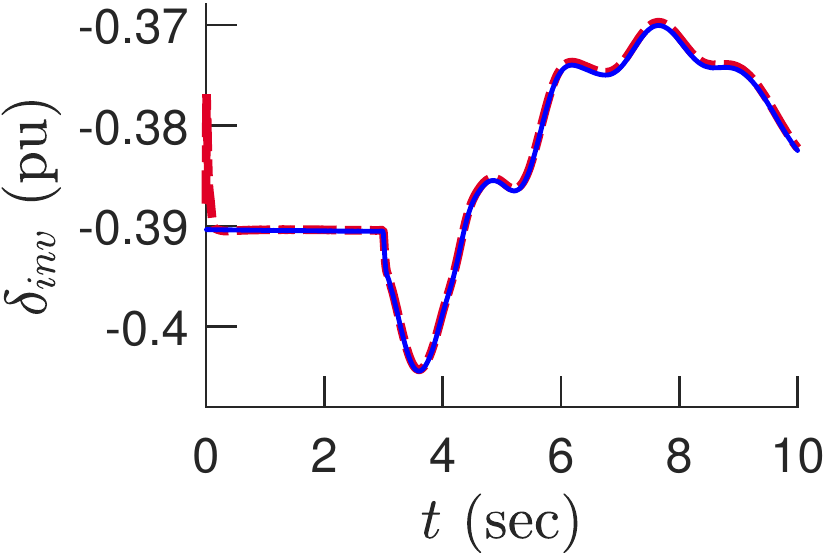}}{}{}\vspace{0.19cm}\hspace{0.2cm}
	\subfloat{\includegraphics[keepaspectratio=true,scale=0.42]{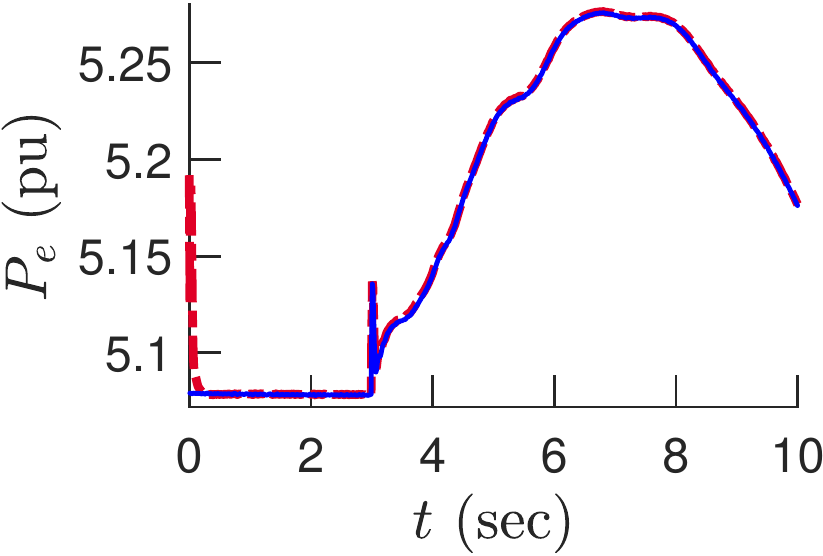}}{}{} \caption{State estimation results for PV power plant 1 under disturbances in real power demand.}\label{fig:estimation PV1}
 \vspace{-0.5cm}
\end{figure}
\begin{figure}
	\centering
	\hspace{-0.19cm}\subfloat{\includegraphics[keepaspectratio=true,scale=0.42]{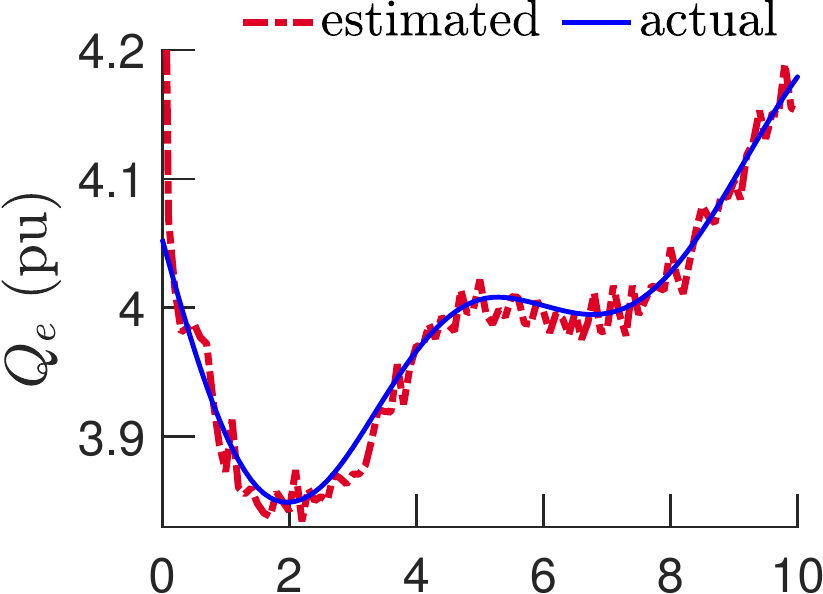}}{}{}\hspace{-0.15cm}
	\subfloat{\includegraphics[keepaspectratio=true,scale=0.42]{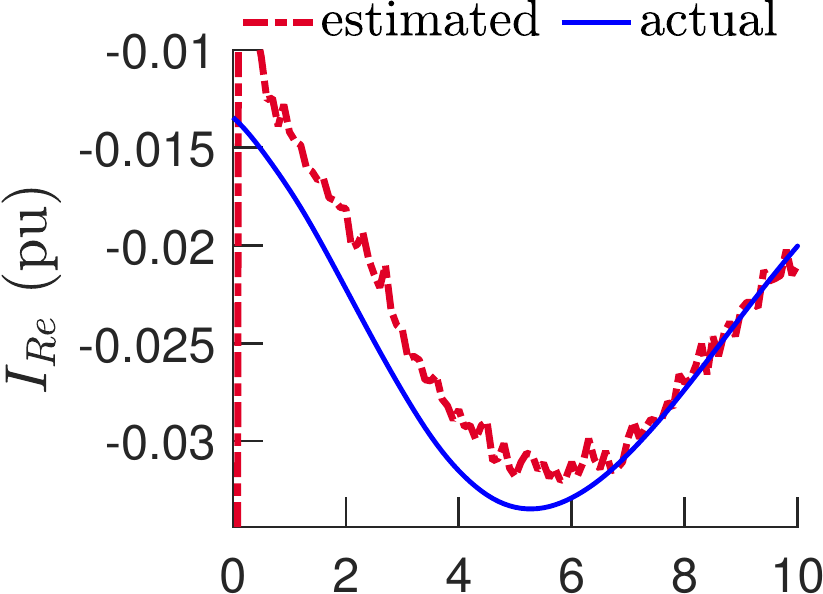}}{}{}\vspace{0.19cm}
	\subfloat{\includegraphics[keepaspectratio=true,scale=0.42]{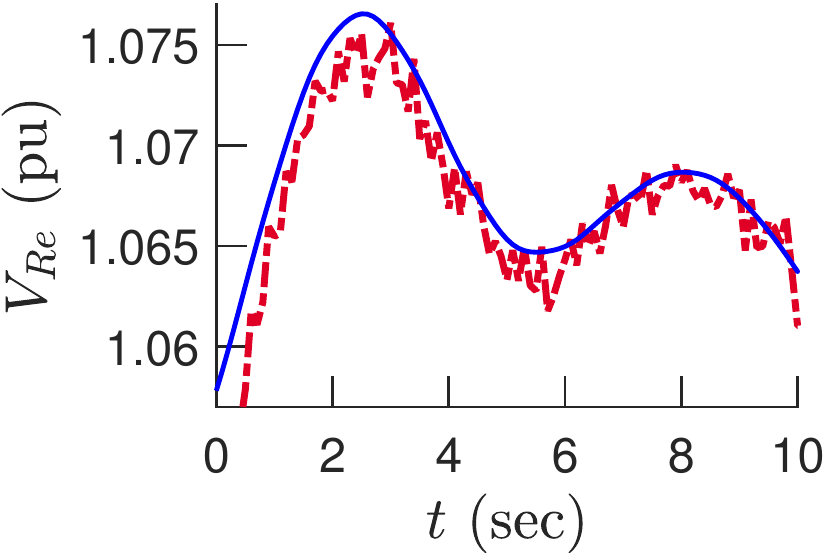}}{}{}\vspace{0.19cm}\hspace{0.2cm}
	\subfloat{\includegraphics[keepaspectratio=true,scale=0.42]{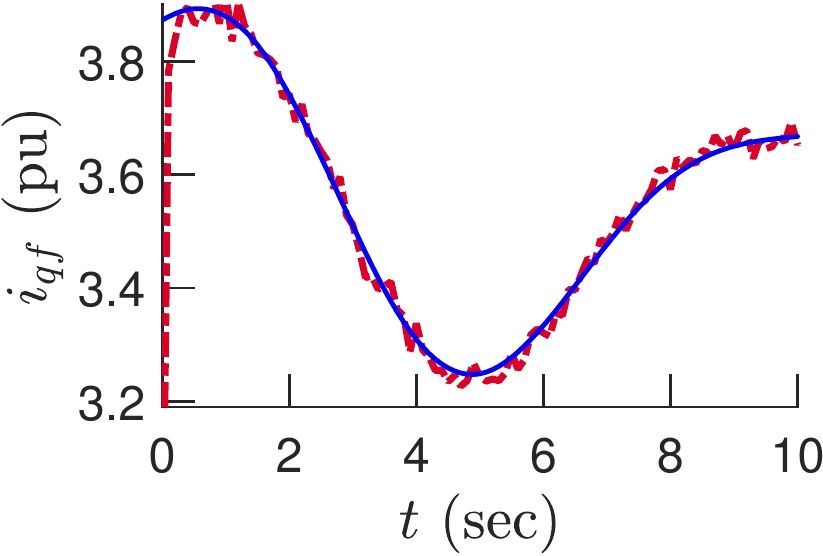}}{}{}
 \caption{Estimation results for PV power plant 2 and algebraic variables of Bus 25 under irradiance and reactive power disturbances.}\label{fig:estimation PV3}
  \vspace{-0.5cm}
\end{figure}

\begin{figure}
\vspace{-0.6cm}
	\centering
	\subfloat{\includegraphics[keepaspectratio=true,scale=0.45]{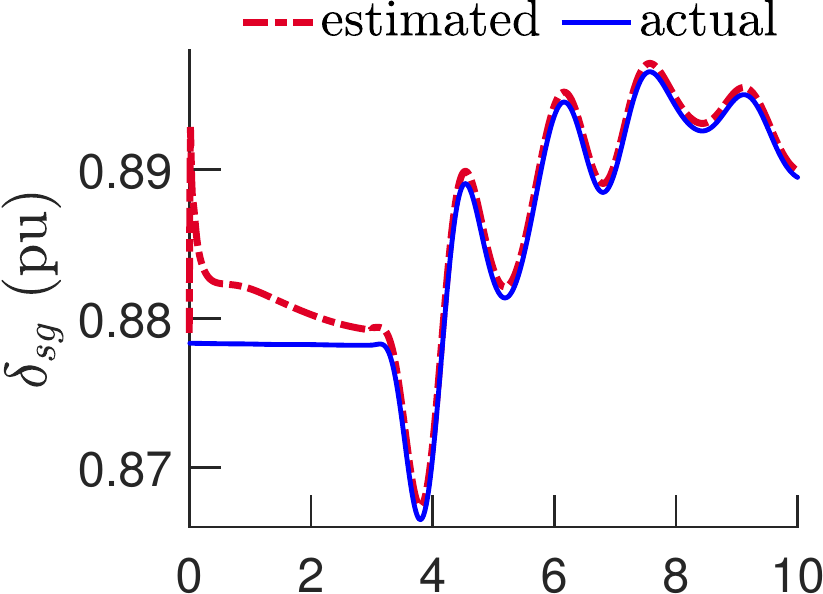}}{}{}\hspace{-0.1cm}
	\subfloat{\includegraphics[keepaspectratio=true,scale=0.45]{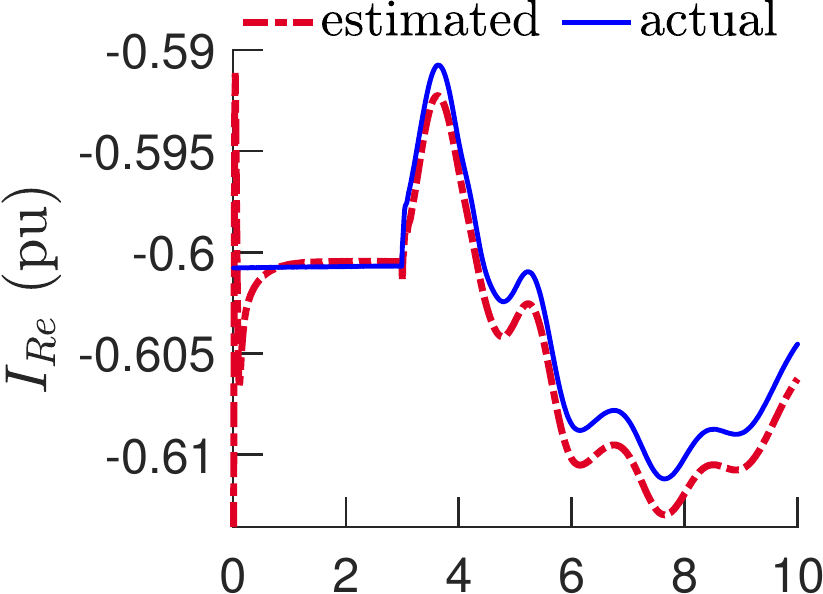}}{}{}\vspace{0.19cm}
	\subfloat{\includegraphics[keepaspectratio=true,scale=0.45]{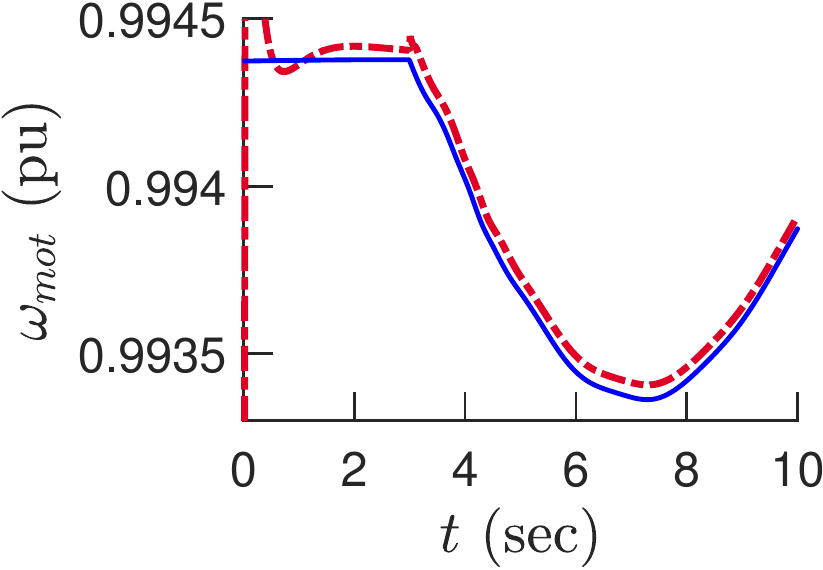}}{}{}\hspace{0.3cm}
	\subfloat{\includegraphics[keepaspectratio=true,scale=0.45]{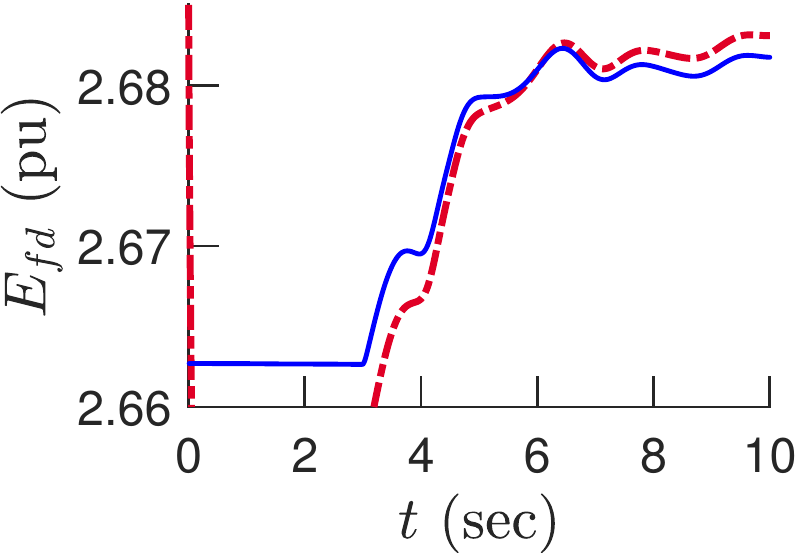}}{}{}\vspace{-0.42cm}
	\caption{Estimation results for Generator-6, real current at Bus 21, and motor speed under disturbance in real power demand.}\label{fig:estimation gen and algebric vars}
 \vspace{-0.5cm}
\end{figure}

The estimation results for the PV power plant 1 are shown in Fig. \ref{fig:estimation PV1}.  We notice that although the estimator starts from initial conditions different than system steady state values and after $t = 3$s the fluctuations in load demand are not known to the estimator, it can still accurately track the different states of the PV power plant. This can also be corroborated from the estimation error norm given in Fig. \ref{fig:Pd and error} from which we can validate that the estimator is successfully driving the error between true and estimated values near zero for all of the system dynamic and algebraic states, under uncertainty. Similar performance from the estimator has been achieved in predicting the internal states of synchronous machines and algebraic variables of all the buses as presented in Fig. \ref{fig:estimation gen and algebric vars}.
\subsection{Case 2: Estimation under irradiance \& power uncertainty}
To further assess the robustness of the estimator toward different sources of uncertainty, in this section we demonstrate the performance of the estimator under disturbances from renewables. To that end, right after $t>0$ we decreased the irradiance from the sun on both of the PV power plants by $20\%$. Also, to create further unknown transient conditions we added step disturbances in the overall reactive power demand of the system. Note that, the estimator is completely unaware of these disturbances and only knows the steady state value of these quantities. This can be validated from the structure of the proposed estimator design \eqref{eq:obsr dynamics}. We can see that the estimator only has access to ${\bar{\m w}}$ which contains the steady state/predicted values of load demand and irradiance.

\begin{figure}
	\centering
	\hspace{-0.4cm}\subfloat{\includegraphics[keepaspectratio=true,scale=0.45]{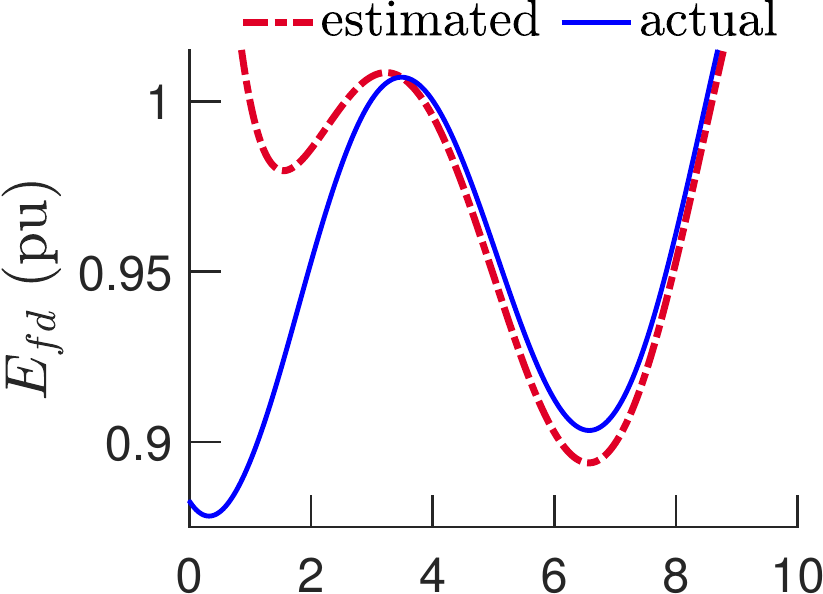}}{}{}\hspace{-0.1cm}\vspace{0.19cm}
	\subfloat{\includegraphics[keepaspectratio=true,scale=0.45]{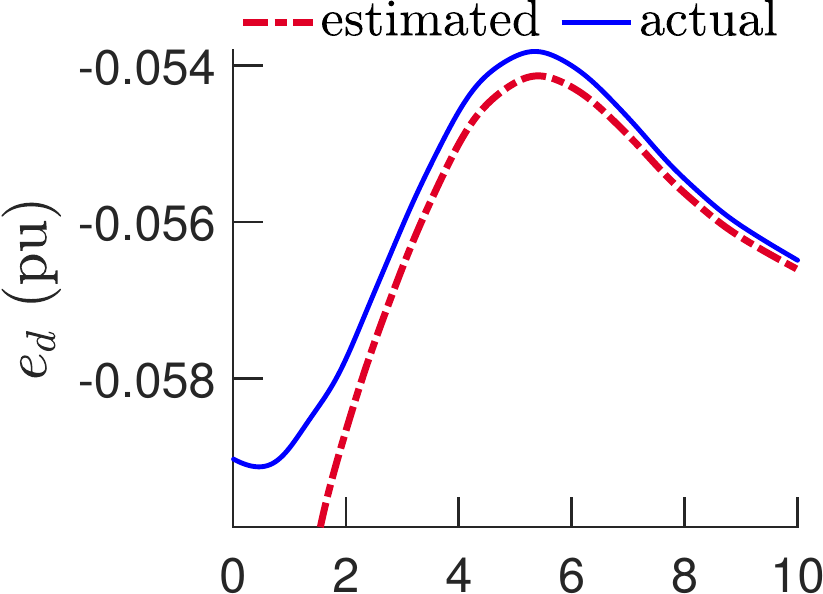}}{}{}
	\subfloat{\includegraphics[keepaspectratio=true,scale=0.45]{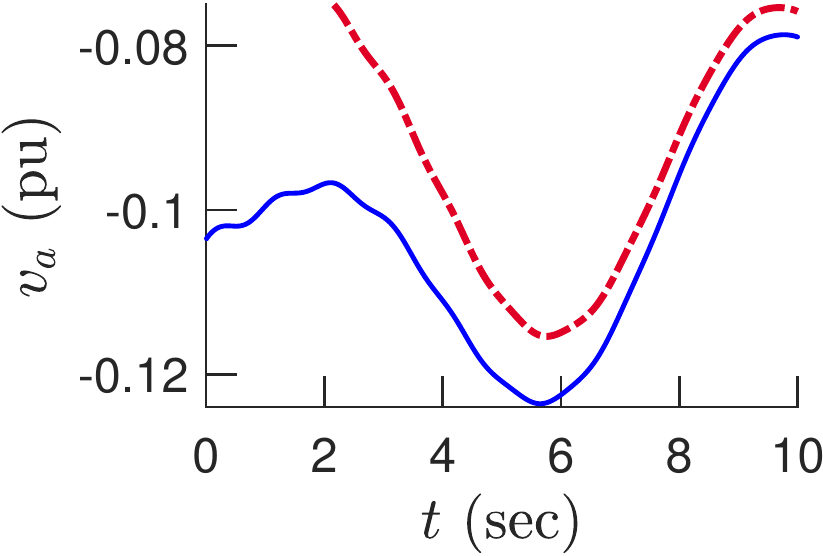}}{}{}\vspace{0.19cm}
	\subfloat{\includegraphics[keepaspectratio=true,scale=0.45]{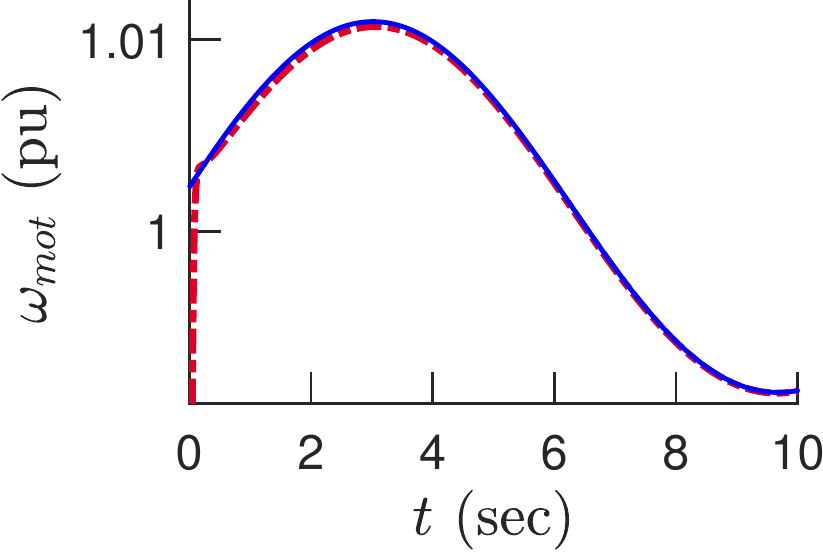}}{}{}\vspace{0.19cm}\hspace{0.2cm}
 \vspace{-0.5cm}
	\caption{Estimation results for Case 2: States of Generator-1 and motor speed.}
 \vspace{-0.55cm}
	\label{fig:estimation Qd and Ir}
\end{figure}

The state estimation results are depicted in Fig. \ref{fig:estimation PV3} and \ref{fig:estimation Qd and Ir}. For brevity estimation results for only a few of the states are shown. We notice that the estimator is still able to track the true values of state variables with good accuracy and is also driving the error norm near zero as shown in Fig. \ref{fig:Pd and error}.

\section{Conclusions and Future Work} \label{sec:conclusion}

In this paper, we proposed a robust estimator for an interconnected model of power systems. The observer design is posed as a convex optimization and work as a one step predictor. Using a few measurements provided by PMUs the proposed estimator can provide an accurate estimate of all the states of a power system including dynamic states of PV power plants and motor loads. The presented estimator does not require any prior knowledge about the statistical properties of the uncertainty and can provide accurate estimation results as long as the uncertainty/disturbance is bounded. 

The limitations of this study are twofold: First, the proposed estimator performs DSE in a centralized fashion, thus for a very big power network it needs to be extended to a decentralized framework. Second, the theory of the proposed estimator is based on continuous models, however PMUs measurements are commonly transmitted via a digitized network, thus a discrete time version of this estimator will be more appropriate. To that end, as a future work the proposed estimator will be extended to a discretized and decentralized framework.

\vspace{-0.5cm}

\section{Acknowledgments}

The power system model used in this study was developed by Dr. Hugo Villegas Pico (hvillega@iastate.edu) and Soummyar Roy (soummyar@iastate.edu)  at Iowa State University. This work is supported by National Science Foundation under Grants 2152450 and 2151571.

\bibliographystyle{IEEEtran} 
\bibliography{mybibfile}


\end{document}